# Ideal Desalination through Graphyne-4 Membrane: Nanopores for Quantized Water Transport

Chongqin Zhu[1], Hui Li[1], Xiao Cheng Zeng[2]*, Sheng Meng[1]*

[1]Beijing National Laboratory for Condensed Matter Physics, and Institute of Physics, Chinese Academy of Sciences, Beijing 100190, China

[2]Department of Chemistry, University of Nebraska-Lincoln, Lincoln, NE 68588, USA

Email: smeng@iphy.ac.cn; xzeng1@unl.edu

## Abstract

Graphyne-4 sheet exhibits promising potential for nanoscale desalination to achieve both high water permeability and salt rejection rate. Extensive molecular dynamics simulations on pore-size effects suggest that γ-graphyne-4, with 4 acetylene bonds between two adjacent phenyl rings, has the best performance with 100% salt rejection and an unprecedented water permeability, to our knowledge, of ~13L/cm$^2$/day/MPa, about 10 times higher than the state-of-the-art nanoporous graphene reported previously (*Nano Lett.s* **2012,** *12*, 3602-3608). In addition, the membrane entails very low energy consumption for producing 1m$^3$ of fresh water, i.e., 3.6×10$^{-3}$ kWh/m$^3$, *three orders* of magnitude less than the prevailing commercial membranes based on reverse osmosis. Water flow rate across the graphyne-4 sheet exhibits intriguing nonlinear dependence on the pore size owing to the quantized nature of water flow at the nanoscale. Such novel transport behavior has important implications to the design of highly effective and efficient desalination membranes.

# Introduction

Although water is abundant on earth, 98% of the available water resource is in the form of salty water.[1] In fact, the shortage of clean and fresh water is one of most pervasive problems afflicting human being's life in the world. The situation deteriorates with increasing human population reaching over 7.1 billion and ever increasing industrialization. Desalination is one viable solution to produce fresh water from salty water. Conventional desalination methods, including reverse osmosis (RO) and thermal desalination, encounter two major obstacles: high energy consumption and extremely expensive infrastructure.[2,3] Commercial desalination based RO requires vast amount of energy at the level of 3.6-5.7 kWh/m$^3$.[2-4] In 2009, production of 1 m$^3$ fresh water cost approximately US $0.10-$1.50. Thermal-based methods such as multi-stage flash and multi-effect distillation are even more inefficient in energy usage.[2,3,5]

Membrane nanoporous materials have attracted considerable attentions due to their great potential for desalination.[4,6-8] Water can flow through very narrow nanopores thanks to its small molecular size (~3Å) while ion passage can be simultaneously blocked given the larger size of ionic hydration shell. For instance, the diameter of sodium hydration shell is ~7.6 Å.[9,10] Two popular nanoporous materials, zeolites[6,7] and carbon nanotube (CNT) arrays[11-13], have been examined for desalination. However, both materials have their disadvantages. First, water flux across zeolite membrane is very low due to complex pore architecture of the zeolite membrane. Although both experiments and molecular dynamics (MD) simulations have demonstrated that CNT can allow fast water flow,[14, 15] low salt molecule/ion rejection along with the difficulty of producing high quality CNT array seriously limit its wide application.[2,4] Graphene monolayer containing artificial nanoscale pores has also been proposed[8]. However, carving nanoscale pores at high density on a graphene sheet remains a formidable technology challenge, awaiting for breakthroughs in experiment and nanomanufacturing.

Graphyne is a new family of carbon allotrope and it is also one-atom-thick planar sheet but with many forms of extended conjugation between acetylene and phenyl groups.[16-20] One form called γ-graphyne, whose structure is built upon phenyl rings fully connected by acetylene bonds, exhibits triangular pores (Fig. 1). The pore size can be adjusted by changing the number of acetylene bonds (defined as $n$) between adjacent phenyl rings. For $n = 0$ it becomes the graphene. Other γ-graphyne structures are referred to as the graphyne-$n$ hereafter. Note that graphyne-1 (or graphyne in the literature) and graphyne-2 (also referred to as graphdiyne) has been successfully produced in large quantities.[18,19] Previous studies have shown that graphyne sheets exhibit novel electronic properties such as high electron mobility and direction-dependent Dirac cone[21,22]. Another first-principles study suggested that graphyne-2 (or graphdiyne) may be used for precise hydrogen purification.[23]

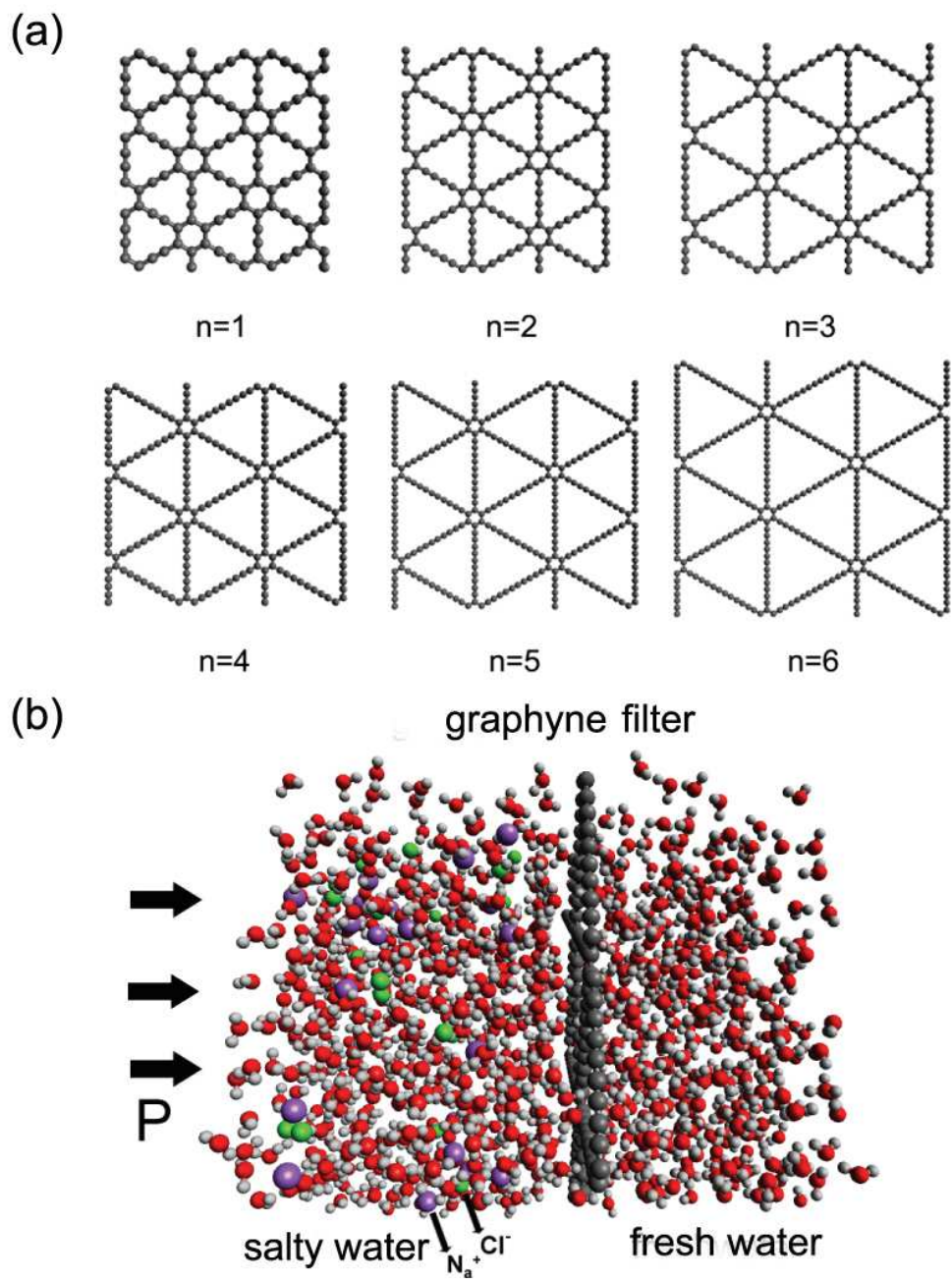

**Figure 1**. (a) Graphyne-$n$ sheets with pore sizes of $n$ =1~6, and (b) a side view of the simulation system. Color codes: C (grey), O (red), H (white), Na$^+$ (purple), and Cl$^-$ (green).

Here, we demonstrate through extensive MD simulations that γ-graphyne with specific nanoscale pores can be an outstanding membrane for water desalination at high rate and potentially low cost. Water can pass through the graphyne membrane when the number of connecting acetylene bonds is no less than three ($n \geq 3$), while significant salt rejection is found for the graphyne membrane with $n \leq 6$. Nearly perfect ion rejection occurs for $n = 4$, with which an unprecedented water flux of ~13L/cm$^2$/day/MPa is achieved, 3 orders of magnitude higher than the commercial RO membranes[4] and ~10 times higher than nanoporous graphene reported previously (even at a very high pore density of 1 nm$^{-2}$)[8]. Moreover, contrary to the optimal porous graphene devices[8], we find that desalination efficiency of the graphyne membrane increases with the applied hydrostatic pressure, suggesting that the desalination can proceed in high speed without at the expense of losing high efficiency. The remarkable performance of the graphyne membrane is attributed to a unique feature of fluid dynamics, namely, the quantized flow of water molecules across dense nanopore array of the graphyne sheet. A better understanding of this feature will allow us to control water fluid dynamics at the nanoscale.

## Computation Methods

**Classical Molecular Dynamics**. Atomic structures of the graphyne-*n* sheet ($n = 1$-6) are shown in Fig. 1a, with parameters obtained from Ref. 24 (side length of nanopores being 6.95 Å, 9.51 Å, 12.07 Å, 14.63 Å, 17.18 Å, and 19.74 Å for $n = 1$~6, respectively). Simulated solution contains 20 Na$^+$ ions, 20 Cl$^-$ ions, and 900 water molecules in a 40×40×150 Å³ box, corresponding to a salt concentration of 6.7%. The higher concentration than natural seawater (2.8%) is to facilitate MD simulations. The graphyne membrane is fixed at the vertical position $z = 0$, with all solution molecules initially placed on one side of the graphyne ($z < 0$). Periodic boundary conditions are applied along in-plane directions. A rigid piston initially placed at $z = -75$Å slowly pushes the solution toward the membrane at a prescribed external pressure (ranging from 100 to 500 MPa). Water flux and the applied pressure show a linear relationship in our simulation (Supporting Information Fig. S1), suggesting that our simulation results remain physical at low pressures, even though the pressure values in MD simulations are higher than typical pressure used in commercial desalination (a few MPa).[2,3]

Non-bonding interactions are modeled by the Lennard-Jones (LJ) and Coulomb potentials. The rigid SPC/E model is employed for water. The interaction parameters for graphyne,[25] $Na^+$ and $Cl^-$ ions are adopted from the Amber99 force field. Particle-mesh Ewald summation and atomic cutoff methods are applied to calculate electrostatic and van der Waals (vdW) interactions,[26] respectively. The simulation systems are firstly equilibrated in a constant-volume and constant-temperature (*NVT*) ensemble for 200 ps at 300 K with the external piston pressure kept at $P = 0$. Each system is then simulated for another 10 ns by maintaining a constant pressure normal to the piston. A timestep of 1.0 fs is selected, and totally five parallel MD trajectories are collected for each system with random distributions of ions to eliminate the initial-position-dependent errors. All the simulations are performed using the Gromacs

4.4.5 package.[27]

**First-principles Simulations**. First-principles simulations are carried out in the framework of density functional theory (DFT) as implement in the *CP2K* package.[28,29] The Perdew-Burke-Ernzerhof (PBE)[30] functional is selected, combined with an empirical dispersion correction for better description of intermolecular interactions.[31] The Goedecker, Teter, and Hutter's (GTH)[32,33] norm-conserving pseudopotential is employed to mimic the interaction between nuclei and core electrons and valence electrons. The GTH-DZVP Gaussian basis along with a plane-wave basis set (energy cutoff of 280 Ry) is applied. A sandwich-like model in a $12.01 \times 12.01 \times 40$ Å$^3$ box is used (Fig. 6a), where a graphyne-3 sheet (with two adjacent trianglar pores per unit cell) is located in the center of the box as the desalination membrane. The piston is located at the bottom of simulation box. Here, the solution contains 78 water molecules and a pair of K$^+$ and Cl$^-$ ions. The system is equilibrated first in an *NVT* ensemble for 40 ps at 360 K with timestep of 1.0 fs. Next, opposite forces are introduced on the piston and graphyne membrane in order to produce 500 MPa external pressure. An additional 20 ps simulation is collected for the study of water desalination.

## Results

**A. Water permeability**

Over 10-ns MD simulation, no event of water permeation across either graphyne-1

or graphyne-2 membrane is observed as the size of nanopores is too small to allow a water molecule pass through. Water starts to pass through the graphyne-$n$ with $n \geq 3$. Water flux is a constant but it increases when the pore size is enlarged or when the applied pressure via piston increases (Fig. S1). In Fig. 2a, we plot water flow rate $V_s$ across a single nanopore as a function of the applied pressure $P$. Water flow rate increases linearly with the pressure, and this linearity extends to very low pressures $P \leq 5$ MPa in our simulations. However, $V_s$ does not show a linear relation with the pore size. Compared to graphyne-3, $V_s$ increases by 1.3, 1.6 and 3.1 times, respectively, for graphyne-4, graphyne-5, and graphyne-6. Although the nanopore size increases significantly from $n=4$ to $n=5$ (with a change in side length $\Delta L=2.56$ Å), $V_s$ exhibits little change. This behavior is more clearly shown by the single-pore permeability $V_{ms}$, defined as the flow rate per unit pressure per nanopore, as displayed in Fig. 2b. Surprisingly, $V_{ms}$ shows a *stepwise increase* as the pore size increases. For $n = 0, 1, 2$, no water flow is seen, hence $V_{ms}= 0$. Then for $n = 3$, $V_{ms}$ becomes non-zero, i.e., $2.1\times10^{-5}$ ps$^{-1}$MPa$^{-1}$ and for $n = 4$ and 5, $V_{ms} = 4.7\times10^{-5}$ and $5.3\times10^{-5}$ ps$^{-1}$MPa$^{-1}$, and it increases to $8.5\times10^{-5}$ ps$^{-1}$MPa$^{-1}$ for $n = 6$. The stepwise increase in $V_{ms}$ with increasing the pore size demonstrates a unique feature of the "quantized" water flow across the graphyne membranes (more discussion below).

Consequently, the effective water flow per area, $V_e$, across the graphyne-4 is much higher than that across the graphyne-5, and even higher than that across the graphyne-6 (Fig. 2c). This is because the distribution density of nanopores on the graphyne-4 is higher than that on graphyne-5 and graphyne-6 membranes. Thus, a non-monotonic relation between the water permeability and pore size can be understood. Indeed, compared to $n = 3$, $V_{me}$ increases by 1.5, 1.1 and 1.5 times, respectively, for $n=4$, 5, and 6 (Fig. 2d). The maximum value of $V_{me} = 13.1$ L/cm$^2$/day/MPa is achieved for the graphyne-4 membrane, *three orders higher* than the current commercial RO membrane ($2.6\times10^{-2}$ L/cm$^2$/day/MPa).[4,34]

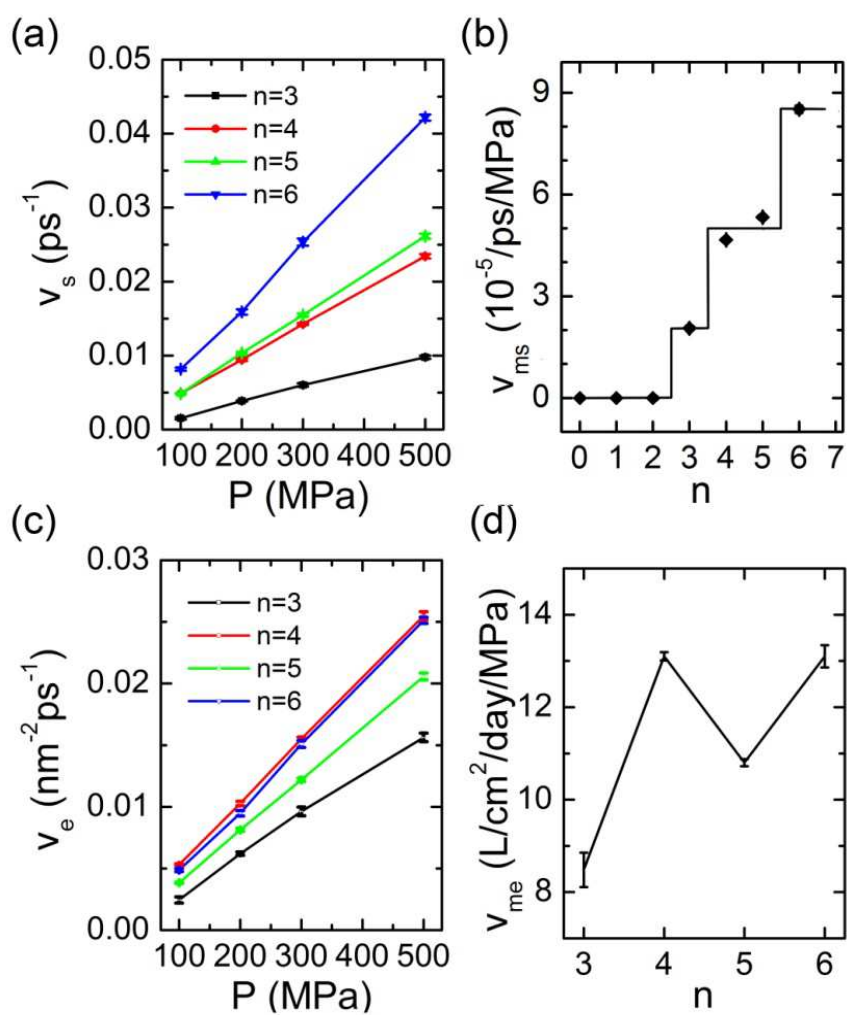

**Figure 2**. (a) Flow rate per nanopore as a function of applied hydrostatic pressure, (b) permeability per nanopore, (c) effective flow rate per area as a function of pressure, and (d) effective permeability per area for graphyne-*n* membrane (*n*=3 to *n*=6). Error bars for each data point are also shown.

**B. Salt rejection**

High salt rejection is observed for all graphyne membranes considered in this study (Fig. 3). The efficiency of salt rejection can be defined as $R = 1-N_{1/2}/N_0$, where $N_0$ is the initial number of ions in the solution and $N_{1/2}$ is the number of ions in permeating solution when half amount of the solution passes through the membrane.[8] Graphyne-3 and graphyne-4 have a perfect salt rejection efficiency ($R = 100\%$), while graphyne-5 has slightly decreased efficiency $R \sim 95\%$ and graphyne-6 maintains a high $R$ value about 80%. As a comparison, the nanotube membranes typically yield much lower salt rejection rates (e.g., ~20%),[13] and even the porous graphenes at high pressure yield lower rejection rates.[13]

Note also that the salt rejection efficiency increases slightly with the applied external pressure. This trend can be explained by the computed energy barrier for passing ion, which is much higher than that for water passing through nanopores. When the pressure is increased, water flux flow through the nanopore increases while ion flux is essentially unchanged. Statistical analysis shows that under higher pressure, more hydration bonds need to break when ions passing through graphyne membranes so that the ion passage becomes more difficult (see Fig. S2). The increased salt rejection under high pressure is consistent with the behavior with the diffusive RO membranes[35], but is contrary to that of the porous graphene, where the pressure increase can result in a decrease in salt rejection efficiency[35]. This striking difference between the graphyne and graphene nanopores might come from that flat edges of graphene nanopore help ion dehydration under high pressure.

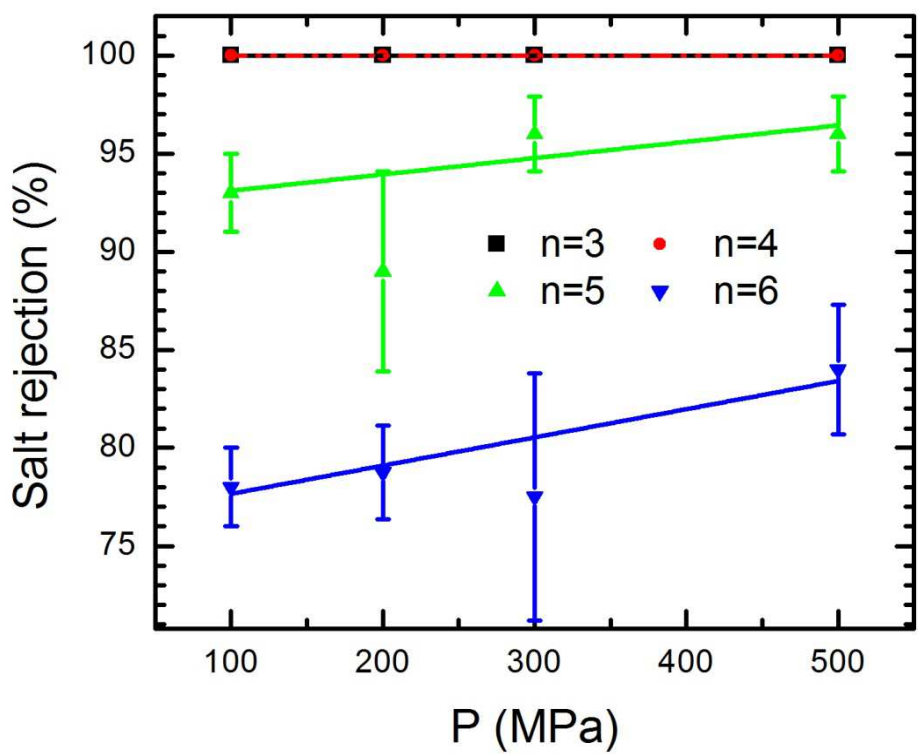

**Figure 3**. Salt rejection efficiencies (with error bars) as a function of applied pressure.

## C. Microstructure of water flow

Simultaneous high water permeation and salt rejection for the graphyne membrane can be attributed to the pore-size effects on the microscopic structure of water flow. Density profile analysis shows that a strong molecular layering occurs close to the graphyne membrane (Fig. S3). The density profile peak at $z = -3.2$ Å is higher than that at $z = +3.2$ Å, indicating a stronger layering on the feeding reservoir resulted from the higher hydrostatic pressure. The high stability of vicinal water layers blocks easy passage of water through the graphyne, which must break well-ordered in-plane hydrogen bond (HB) network.

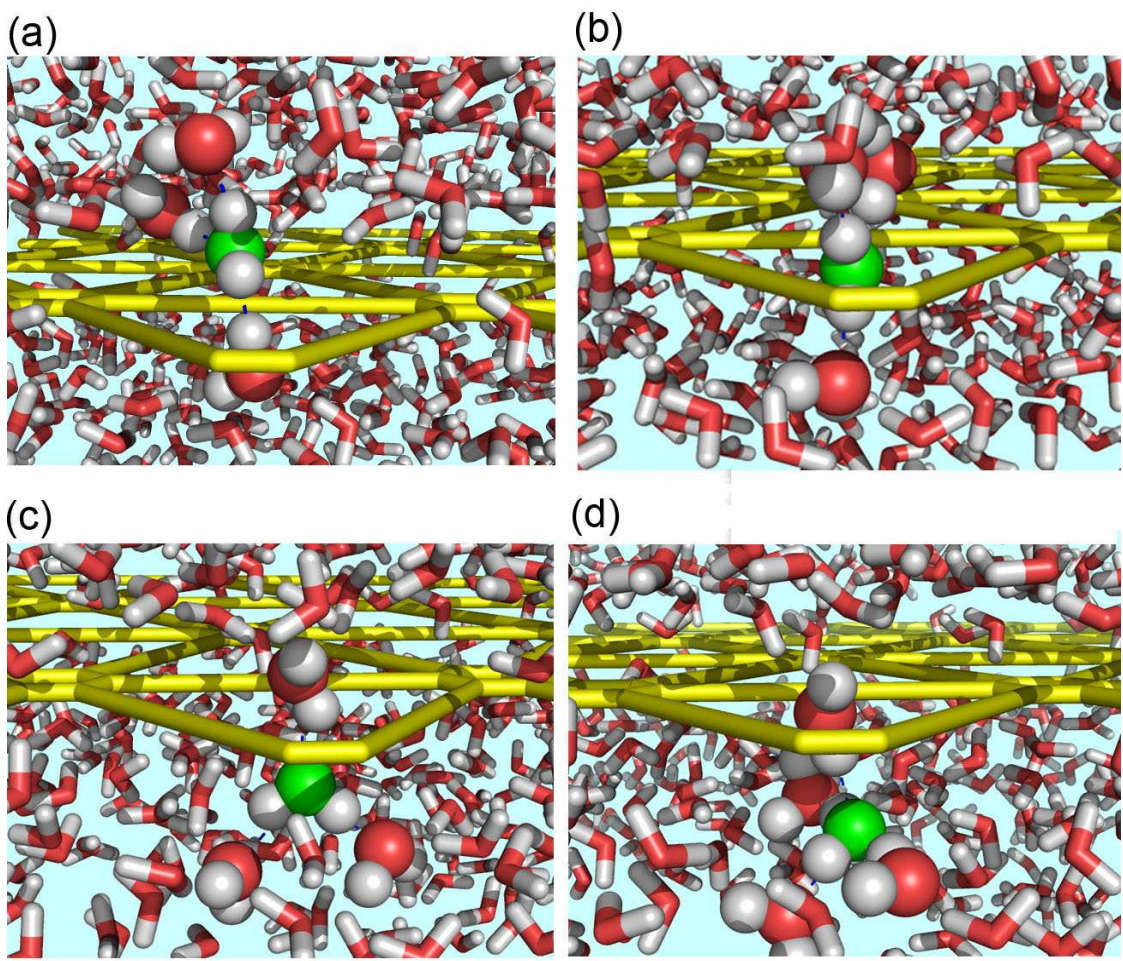

**Figure 4**. Snapshots of a water molecule (spheric model, oxygen in green color) transporting through graphyne-3 membrane. Water molecules forming hydrogen bonds with this water are shown in spherical models with oxygen in red. Graphyne sheet is shown in yellow and other water molecules in stick model.

To gain insights into the process of water passing through the membrane, we analyze a series of snapshots when water passing through graphyne-3 sheet, as shown in Fig. 4. A water molecule at the center of a triangular pore first breaks its in-plane hydrogen bond (HB), and adopts a configuration with its molecular plane perpendicular to the graphyne surface. Next, it forms HBs with water molecules on both sides of the membrane, a transition state before passing through the membrane (Fig. 4b). As soon as the water molecule enters the nanopore and arrives in the other side, it rapidly rotates itself to form three HBs: two are connected to the in-plane network and the third is connected to the next incoming water molecule for passage (Fig. 4c). At the end of passing process, the water becomes a member of the structured layer on the fresh-water side of the membrane (Fig. 4d). Unlike the continuum macrofluid, the nanofluid passing through graphyne membrane can be viewed as taking a quantized manner, passing water molecular one at a time, which results in strong dependence of water permeability on the nanopore size.

The pore-size effect can be clearly illustrated in the 2D density maps of O and H atoms in water (Fig. 5). The graphyne network has a strong affinity to water. Only one maximum of O-density and three maxima of H-density are observed inside the nanopores of the graphyne-3, corresponding to three equally probable orientations of a passing water molecule. Density maps for graphyne-4 and graphyne-5 are similar to one another: three major maxima can be seen near the vertex regions, illustrating that water prefer to pass through those regions. Although the pore size in graphyne-5 is much larger than that in graphyne-4, most area is ineffective for water transport, indicated by the large blue regions in the O-density map of graphyne-5. Three additional subtle density maxima nearby the triangle sides start to form and become more intensified for graphyne-6, indicating that the three sides of nanopore become more important for water passage through large nanopores.

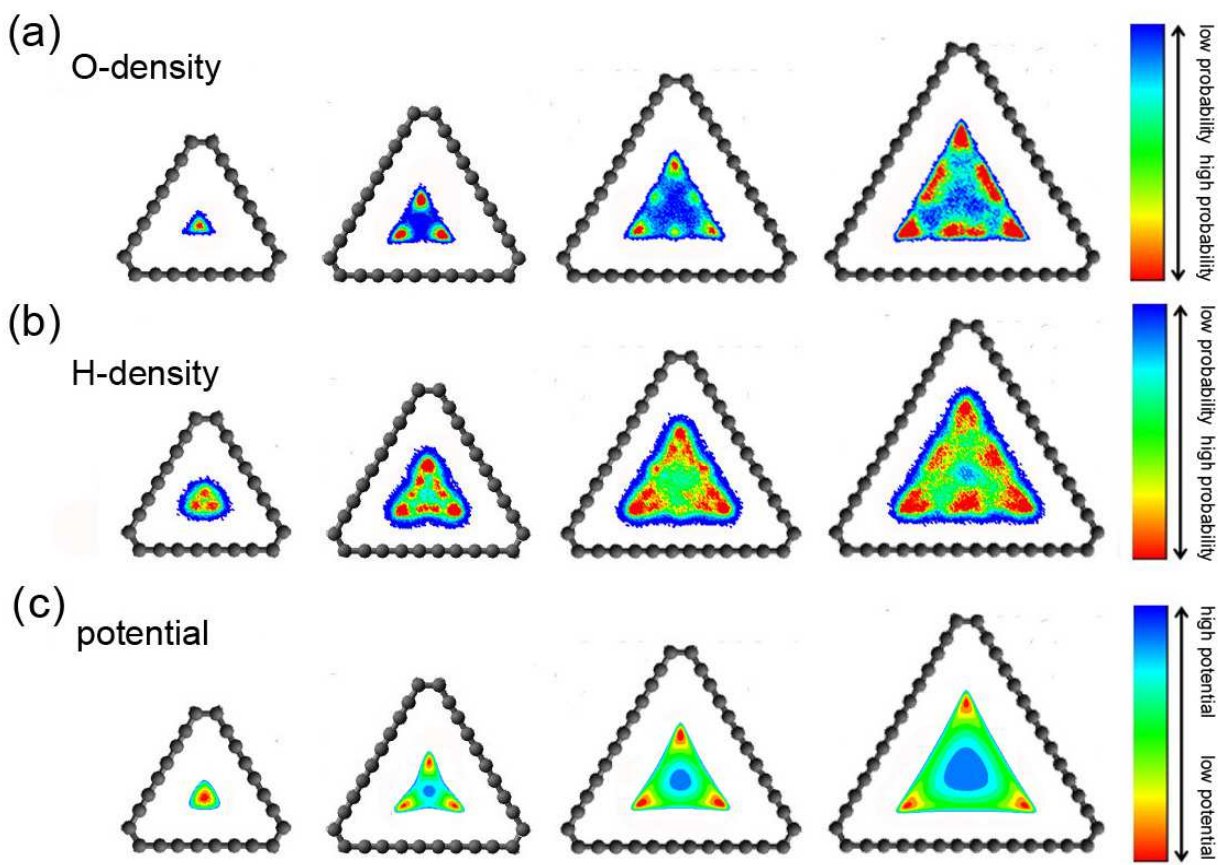

**Figure 5**. (a) Oxygen density distribution and (b) hydrogen density distribution inside a nanopore of graphyne-*n* (*n*=3 to *n*=6). (c) Potential energy of a water molecule inside the nanopore of graphyne-*n* membrane.

# Discussion

Discrete O and H density distributions shown in Fig. 5 stem from water-graphyne interaction, beaconed by the potential energy of water inside the nanopores as shown in Fig. 5c. Only one minimum is observed in the water potential map for graphyne-3, suggesting only a single water molecule can pass through at a time. This is confirmed by the statistics on the number of synchronously passing water molecules, which is exactly one (Fig. S4a). Distribution of water orientation, characterized by the angle $\theta$ between the OH bond of water and $z$-axis normal to graphyne membrane, displays two major peaks at $\theta=45º$ and $125º$ for graphyne-3 (Fig. S4b), indicating that the passing water must adopt an appropriate direction with two OH bonds symmetrically pointing to water layers on both sides of the membrane.

Potential maps of graphyne-$n$ ($n=4$ - 6) show three minimum spots in the vertex regions. Almost all water molecules pass through these three spots for graphyne-4 and 5 with relative smaller pores. Water can be pulled through side positions of the triangles by those at the vertex corners (potential minima) for graphyne-6 with relatively larger pores, where three additional density maxima show up near the triangular sides. Graphyne-5 has the same number of effective water pathways as graphyne-4 due to steric hindrance, despite of appreciable increase in pore side length $\Delta L = 2.56$ Å. This is the reason why their single-pore water permeability $V_s$ is nearly the same. We also note that the number distribution of instantaneous water fluxes is almost identical for graphyne-4 and 5 (Fig. S4a), in spite of relatively high probability (> 25%) of double-molecule flow.

Water flux shows a similar angular distribution for graphyne-4 and graphyne-6 with three peaks at $\theta=35º$, $85º$ and $140º$ (Fig. S4b), respectively, implying water molecules adopt either vertical or parallel orientation with respective to the graphyne plane during the passing. Water requires less reorientation energy as it passes through graphyne-3, hence much increase of water flow. Surprisingly, there is no obvious peak in water orientation distribution for graphyne-5, implying no special orientation is required. Compared to graphyne-4 and graphyne-6, graphyne-5 represents a transition from "quantum" steps of graphyne-4 to graphyne-6 where water flow is floppy before additional water can pass through and new flow patterns being established.

When the pore size is large enough (graphyne-6), steric hindrance decreases and nanopore sides become important for water permeability. The number of synchronously passing molecules varies from one to four with a Boltzmann-like distribution centered at two, indicating the size effect on quantized water flux is smeared. As a comparison, water flowing through a narrow CNT can be in a form of 1D single-file chain, or even exhibit a subcontinuum to continuum flow transition at a pore diameter of ~13 Å.[36,37,37b] The quantized nature of water flow across the graphyne membrane ($n$=3 - 6) offers a physical explanation of the apparent steps of water flow as a function of pore size (Fig. 2b).

Lastly, we perform the first-principles MD (FP-MD) simulations to validate empirical parameters used in classical MD. Our FP-MD simulations show that the graphyne sheets exhibit very strong mechanical strength sustaining high hydrostatic pressure (500 MPa) and relatively long FP-MD simulation time (60 ps) without noticeable deformation (Fig. 6). The in-plane HB network starts to break at 16 ps for graphyne-3 under 500 MPa, and the water pass through the membrane in the same manner as observed from classical MD simulation (Fig. 6c), giving an estimated water flow rate of $5 \times 10^{-5}$ /ps/MPa (the same order as from classical MD). More importantly, both cations and anions stay away from the graphyne membrane during entire FP-MD simulation, suggesting perfect salt rejection.

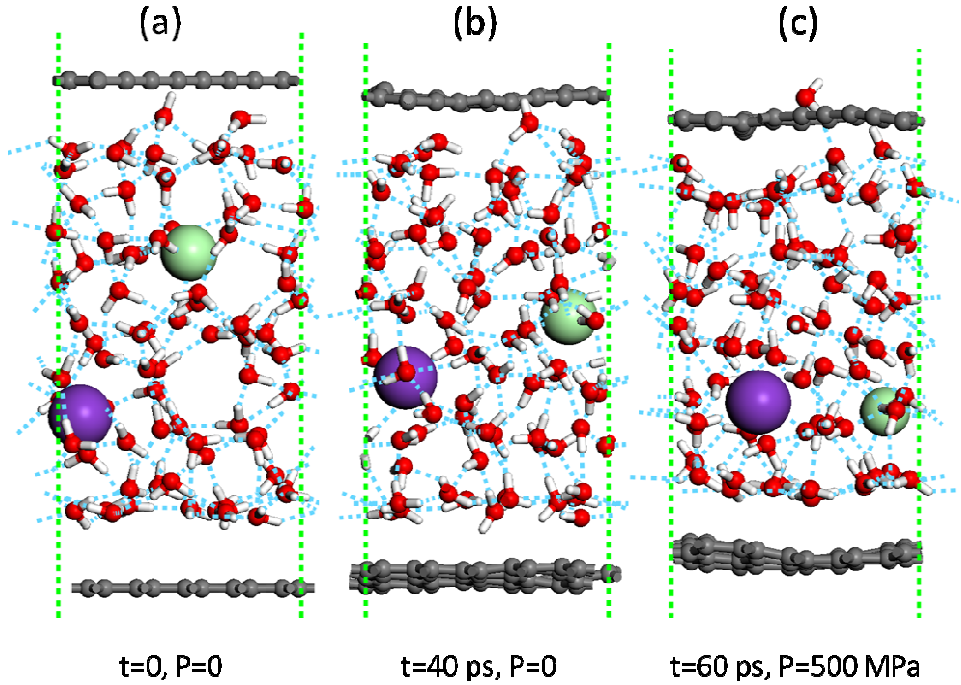

t=0, P=0    t=40 ps, P=0    t=60 ps, P=500 MPa

**Figure 6**. Snapshots of salty water confined by graphene (bottom) and graphyne-3 (top) membranes at various time of first-principles molecular dynamics simulation: (a) t = 0 (P=0), (b) t = 40 ps (P=0), and (c) t = 60 ps (P=500 MPa). Color codes: C, gray; O, red; H, white; Alkaline metal, purple; Cl, green; H-bond, blue dash line.

# Conclusion

A conventional notion in the field of desalination is that high salt rejection and high water flux across membranes are two competing factors. Our MD simulations show that the 2D graphyne sheet can achieve both high water permeability and high efficiency for salt rejection at the same time, thereby a promising candidate as ideal desalination membranes. In particular, we show that the graphyne-4 can achieve the highest theoretical water flux of ~13L/cm$^2$/day/MPa along with 100% salt rejection, as well as three orders of magnitude lower in energy consumption ($3.6\times10^{-3}$ kWh/m$^3$) compared to the state-of-art commercial RO membranes (1.8 kWh/m$^3$). Finally, we note that some forms of graphyne and its derivatives have been synthesized in the laboratory,[18,19,38-43] and graphyne-1 and graphyne-2 (graphdiyne) have been successfully produced in large quantities.[18,39,40,44] It is expected that synthesis of graphyne-4, graphyne-5, and graphyne-6 with relatively larger pores may be achieved in near future, given the intensive current interests and efforts to synthesize all-carbon conjugated networks.

## Acknowledgements

We acknowledge financial support from NSFC (grants 11074287 and 11222431), MOST (grant 2012CB921403), water-projects cluster and hundred-talent program of CAS. XCZ was supported by grants from the NSF (CBET-1036171 and CBET-1066947) and ARL (W911NF1020099).

## Supporting Information

The number of filtered water molecules per area as a function of simulation time, ion coordination number as a function of the distance (height) from the ion to the graphyne-5 membrane, water density along the direction $z$ normal to the graphyne membrane, as well as distribution of the number of water molecules synchronously passing through individual nanopores. This material is available free of charge via the Internet at http://pubs.acs.org.

TOC Graphic

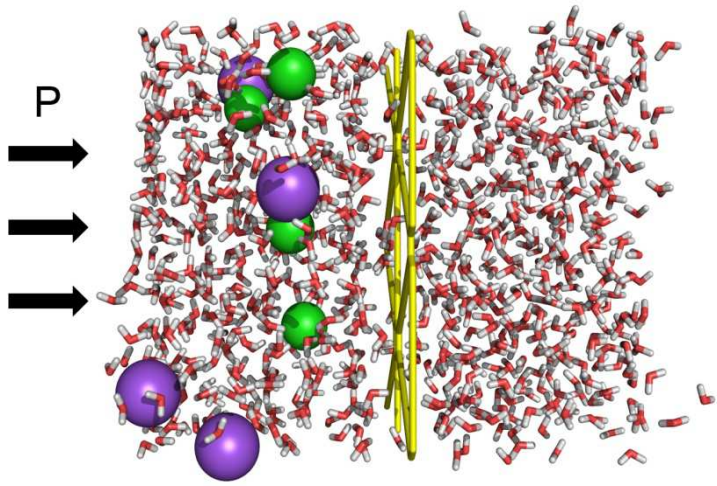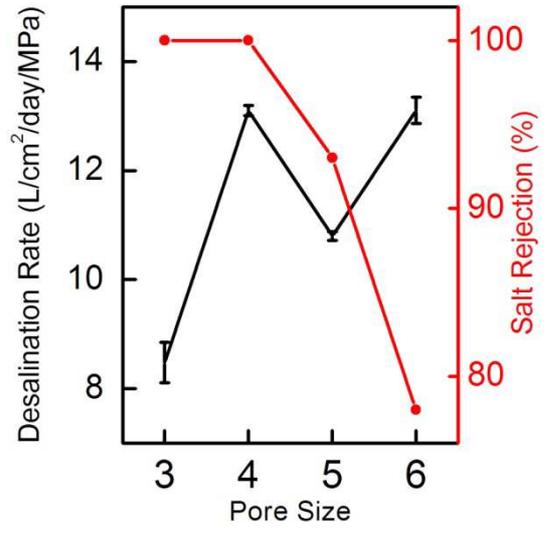